\documentstyle[epsf]{article}
\newskip\humongous \humongous=0pt plus 1000pt minus 1000pt

\newif\ifdtup

\catcode`\@=11

\@addtoreset{equation}{section}
\def\theequation{\arabic{equation}}

\def\@normalsize{\@setsize\normalsize{15pt}\xiipt\@xiipt
\abovedisplayskip 14pt plus3pt minus3pt%
\belowdisplayskip \abovedisplayskip
\abovedisplayshortskip \z@ plus3pt%
\belowdisplayshortskip 7pt plus3.5pt minus0pt}

\def\small{\@setsize\small{13.6pt}\xipt\@xipt
\abovedisplayskip 13pt plus3pt minus3pt%
\belowdisplayskip \abovedisplayskip
\abovedisplayshortskip \z@ plus3pt%
\belowdisplayshortskip 7pt plus3.5pt minus0pt
\def\@listi{\parsep 4.5pt plus 2pt minus 1pt
     \itemsep \parsep
     \topsep 9pt plus 3pt minus 3pt}}

\relax

\catcode`@=12

\catcode`\@=11

\def\section{\@startsection{section}{1}{\z@}{3.5ex plus 1ex minus
   .2ex}{2.3ex plus .2ex}{\large\bf}}

\def\thesection{\arabic{section}.}

\def\appendix{\setcounter{section}{0}
 \def\thesection{Appendix \Alph{section}:}
 \def\theequation{\Alph{section}.\arabic{equation}}}
\begin{document}
\begin{titlepage}
\begin{center}
{\Large
Quark Number  Fractionalization in $N=2$ Supersymmetric\\
$SU(2) \times U(1)^{N_f}$ Gauge Theories
}
\end{center}
\vspace{1em}
\begin{center}
{\large Giuseppe Carlino$^{(1)}$, Kenichi Konishi$^{(1)}$ \\
and \\Haruhiko Terao$^{(2) }$ }
\end{center}
\vspace{1em}
\begin{center}
{\it
Dipartimento di Fisica -- Universit\`a di Genova$^{(1)}$\\
Istituto Nazionale di Fisica Nucleare -- Sezione di Genova$^{(1)}$\\
Via Dodecaneso, 33 -- 16146 Genova (Italy)\\
Department of Physics, Kanazawa University$^{(2)}$ \\
Kanazawa, Japan\\
E-mail: konishi@infn.ge.infn.it; terao@hep.s.kanazawa-u.ac.jp
}
\end{center}
\vspace{3em}
\noindent
{\bf ABSTRACT:}

  {
Physical quark-number charges  of    dyons
are determined, via a formula which generalizes that of Witten for the 
electric charge,    in  $N=2$ supersymmetric  theories  with
$SU(2) \times U(1)^{N_f}  $ gauge group.  The quark numbers of the massless
monopole at a nondegenerate singularity of QMS turn out to vanish
in all cases. A puzzle related to CP invariant cases is solved. 
Generalization of our results to  $SU(N_c)\times U(1)^{N_f}$ gauge theories 
is straightforward.   
\\
\\
\\}

\vspace{1.5em}
\begin{flushleft}
GEF-Th-11/1997;
KANAZAWA 97-21\\
~~~~~~~~~~~~~~~~~~~~~~~~~~~~~~~~~~~~~~~~~~~~~~~~~~~~~~~~
~~~~~~~~~~~~~~~~~~~~~~~
~~~~~~~~~~~~~~~December 1997
\end{flushleft}
\end{titlepage}
%%%%%%%%%%%%%%%% latex definitions
\newcommand{\beq}{\begin{equation}}
\newcommand{\eeq}{\end{equation}}
\newcommand{\bea}{\begin{eqnarray}}
\newcommand{\eea}{\end{eqnarray}}
\newcommand{\beas}{\begin{eqnarray*}}
\newcommand{\eeas}{\end{eqnarray*}}
\newcommand{\defi}{\stackrel{\rm def}{=}}
\newcommand{\non}{\nonumber}
%%%%%%%%%%%%%%%%%%%%%%%%%%%%%%%%%% definitions
\def\dirac{{\cal D}}
\def\dplus{{\cal D_{+}}}
\def\dminus{{\cal D_{-}}}
\def\dbar{\bar{D}}
\def\H{\cal{H}}
\def\de{\partial}
\def\si{\sigma}
\def\sb{{\bar \sigma}}
\def\rn{{\bf R}^n}
\def\r4{{\bf R}^4}
\def\s4{{\bf S}^4}
\def\ker{\hbox{\rm ker}}
\def\dim{\hbox{\rm dim}}
\def\sup{\hbox{\rm sup}}
\def\inf{\hbox{\rm inf}}
\def\infi{\infty}
\def\nrm{\parallel}
\def\nrmi{\parallel_\infty}
\def\om{\Omega}
\def\Tr{ \hbox{\rm Tr}}
\def\const{\hbox {\rm const.}}
\def\o{\over}
\def\th{\theta}
\def\im{\hbox{\rm Im}}
\def\re{\hbox{\rm Re}}
\def\bra{\langle}
\def\ket{\rangle}
\def\Arg{\hbox {\rm Arg}}
\def\Re{\hbox {\rm Re}}
\def\Im{\hbox {\rm Im}}
\def\longvert{{\rule[-2mm]{0.1mm}{7mm}}\,}

%%%%%%%%%%%%%%%%%%%%%%%%%%%%%%%%%%%%%%%%%%%%%
\noindent{\bf Introduction}

The breakthrough achieved in the celebrated works of Seiberg and Witten
\cite{SW1,SW2} has
made
 possible, for the first time, to go beyond the semiclassical quantization in
the study  of soliton dynamics
in   non Abelian gauge theories in  four dimensions.\cite{Solitons} Of
particular interest
among their properties  is the electric and quark-number fractionalization.  In
Ref.\cite{KT}   it was shown  that, in the semiclassical limit,  the
 exact Seiberg-Witten prepotentials  
and  mass formula reproduce these   effects correctly, in accordance  with the
standard semiclassical calculations \cite{Ferrari,Niemi}.
In the case of the electric charge fractionalization   one has here
an exact quantum result valid   even where   the semiclassical approximation
breaks down.

The situation for the  quark-number fractionalization is somewhat different.
As long as the   $U(1)$ symmetries associated with the quark numbers are
global,
the "physical quark number of monopoles" is a somewhat obscure quantity, even
though such a quantum number is conserved and not spontaneously broken. In
fact,  there is no local field within the theory coupled to the 
 conserved quark-number currents
$J_i^{\mu}$.   For instance  the correlation functions,
 \beq   \Pi^{\mu \nu } (Q)= i \int
d^4x  \, e^{-i Q x} \,  \bra T\{ J_i^{\mu}(x) \,J_i^{\nu}(0) \} \ket
\label{cortrel} \eeq
cannot be easily analyzed at low energies,
although at high energies
these   can be computed perturbatively  due
to asymptotic freedom.

In this paper we consider  $N=2$  supersymmetric $SU(2) \times
U(1)^{N_f} $ gauge theories ($N_f=1,2,3$)   where the Abelian  factors
correspond to
the conserved  quark numbers; more precisely we consider these theories  in
the limit
   \beq g_i  \to 0+, \label{weakcoupl}\eeq
$g_i$ being the
$U_i(1)$ coupling constant.  In other words,  we introduce the hypothetical
weak
 $U(1)^{N_f} $    gauge
bosons and their $N=2$ partners in order  to probe the strong interaction
dynamics, which is dominated by the $SU(2)$ interactions.
  This
theory has a large vacuum degeneracy parametrized by $N_f +1$ moduli
parameters,
\beq     u= \bra \Tr \Phi^2 \ket ; \quad a_i = \bra A_i \ket = m_i/\sqrt2, \,\,
(i=1,2,\ldots N_f). \label{vevs}\eeq
The  physical $i$-th quark number (charge) $S_i$ of a given
particle is by definition  its low energy  coupling strength
 to the $U_i(1)$  gauge boson, measured in the unit of the
coupling constant,  $g_i$.  $g_i$ is common to all particles  (elementary
  and solitonic) and depends only on the scale  while  the charge
$S_i$
depends on the particle considered. For an  elementary particle, say
the $j$-th quark,  $S_i=\delta_{ji}.$   The latter is not renormalized, as
is well known.  Our main aim is to determine the value of $S_i$ for a
given dyon, in each  vacuum $(u, a_1, a_2, \ldots,  a_{N_f}).$

\noindent{\bf Fractional quark numbers of dyons as boundary effects}

The theory  is described by the Lagrangian,
\bea
L&=& {1\over 8 \pi} \im \, \tau_{cl} \left[ \int d^4 \th \,
\Phi^{\dagger} e^V \Phi +\int d^2 \th\,{1\o 2} W W \right]
+  \non \\
  &+& \sum_{i=1}^{N_f}  {1\over 8 \pi} \im \, \tau_{i} \left[ \int
d^4 \th \, A_i^{\dagger} A_i  +\int d^2 \th\,{1\o 2} W_i W_i \right]
+ L^{(quarks)},
\label{lagrangian}
\eea
where
\beq
L^{(quarks)}= \sum_i [ \int d^4 \th \, \{ Q_i^{\dagger} e^V Q_i \, e^{V_i}+
{\tilde Q_i}  e^{-V} {\tilde Q}_i^{\dagger} \, e^{-V_i} \}  +
\int d^2 \th  \sqrt{2} \{ {\tilde Q}_i \Phi Q^i +
A_i {\tilde Q}_i  Q^i \} + h.c.],
\eeq
where $\{\Phi, W\}$  and $\{A_i, W_i\}$  are  $N=2$ vector supermultiplets
containing  the gauge bosons.
Classically this theory has the flat directions parametrized by the vevs
Eq.(\ref{vevs}).   In a generic point of such vacuum moduli space, the gauge
symmetry is broken to $U(1) \times
U(1)^{N_f} $  and the low energy effective
Lagrangian must describe  $N_f+1$ massless gauge bosons and their
superpartners, and eventually, also light quarks or  dyons.   Its  form
is restricted
 by  $N=2 $ supersymmetry to be
\beq
L^{(eff)} = {1\over 8 \pi} \im \, \sum_{i,j=0}^{N_f} 
\int d^2 \th\,{1\o 2} \tau_{ij}
W_i W_j + \sum_{i=0}^{N_f} \int d^4 \th
\,{\de F \o \de A_i} {\bar A}_i   + L^{(light)}, \label{lagrangianbis}\eeq
where $ \tau_{ij}= {\de^2 F \o \de A_i \de A_j},  $
and $L^{(light)}$ describes either light quarks or light dyons.
$F(a_0, \{a_i\}; \newline \Lambda; \{\Lambda_i\}) $ is the prepotential,
holomorphic in its arguments. Also,  we introduced
a notation  \beq    A_0 \equiv A, \quad V_0 \equiv V, \qquad  
 A_{D0} \equiv A_D, \quad V_{D0} \equiv V_D,  \eeq
to indicate the vector multiplet related  to the original
$SU(2)$ gauge multiplet.  $ \Lambda_i $ is the
position of the Landau pole associated to  the $i$-th $U(1)$ gauge interaction.

The form of $L^{(light)}$ near  one    of the quark
singularities (when $u=m_i^2 \gg \Lambda^2$),   is fixed since  the
quantum numbers of the
light quarks with respect to the  $U(1) \times
U(1)^{N_f} $  gauge group are  known.

On the other hand, the monopoles acquire the quark numbers dynamically.
Semiclassically  they arise through the zero modes of the  Dirac Hamiltonian
  in the monopole background \cite{JR,Solitons,SW2,KT}.
  The {\it classical}  quark numbers
of the  dyons, which become massless at various singularities of QMS  are
given in Table 1.
\begin{table}{
\leftskip 1cm \rightskip 1cm
{\bf Table 1}: Classical quark number charges and other global
 quantum numbers of
light dyons.  $\pm$ denotes the  $SO(2N_f)$ chirality of the spinor
representation.   $\theta_{eff}$ gives the value of the effective $\theta$
parameter where the corresponding  dyon becomes massless.
}
\bigskip \\
$N_f=1$
 \begin{center}
 \begin{tabular}{|c|c|c|c|c|c|}  \hline
 name &  $n_1$  & $n_m $ & $n_e$ & $SO(2)$ & $\theta_{eff}$ \\ \hline\hline
 $M$ & $0$ & 1 & 0 & $+$ & $0$ \\ \hline
 $M'$ & $1 $ & 1 & 1 & $-$ & $-\pi$\\ \hline
 $M''$ & $0$ & 1 & 2 & $+$ & $-2\pi$\\ \hline
 \end{tabular}
 \vskip .3cm
 \end{center}
$N_f=2$
 \begin{center}
 \begin{tabular}{|c|c|c|c|c|c|c|c| }  \hline
 name&  $n_1$  & $n_2$ & $n_m$ & $n_e$ &  $SO(4)$  & $SU(2)\times
 SU(2)$ & $\theta_{eff}$\\ \hline\hline
 $M_1 $ & $0$ & $0$ & 1 & 0 & $+$
 & (${\bf 2}, \,\, {\bf 1}$) & $0$\\
 $M_2$ & $1 $ & $1 $ & 1 & 0 & $+$
 & (${\bf 2}, \,\, {\bf 1}$) & $0$\\ \hline
 $M'_1 $ & $1 $ & $0$ & 1 & 1 & $-$
 & (${\bf 1}, \,\, {\bf 2}$) & $-\pi$\\
 $M'_2$ & $0$ & $1 $ & 1 & 1 & $-$
 & (${\bf 1},\,\,{\bf 2}$) & $-\pi$\\ \hline
 \end{tabular}
 \vskip .3cm
 \end{center}
$N_f=3$
 \begin{center}
 \begin{tabular}{|c|c|c|c|c|c|c|c|c|}  \hline
 name & $n_1$  & $n_2$ & $n_3$ & $n_m$ & $n_e$ &
$SO(6)$ & $SU(4)$
 & $\theta_{eff}$ \\ \hline\hline
 $M_0$ & $0$ &$0$ &$0$ & 1& 0& $+$  & ${\bf 4}$ & $0$\\
 $M_1$ & $1$ &$1$ &$0$ & 1& 0& $+$ &${\bf 4}$ & $0$\\
 $M_2$ & $1$ &$0$ &$1$ & 1& 0& $+$ & ${\bf 4}$ & $0$\\
 $M_3$ & $0$ &$1$ &$1$ & 1& 0& $+$ & ${\bf 4}$ & $0$\\ \hline
  $N$ & $0$ &$0$ &$0$ & 2& 1&  & ${\bf 1}$ & $-\pi/2$ \\ \hline
 \end{tabular}
 \vskip .3cm
 \end{center}
\end{table}
Note that, in contrast to Ref.\cite{KT} we have chosen 
  the {\it
classical}  quark number charges  for monopoles to start with.
\footnote{This is, strictly speaking, unnecessary. One can  start with 
any choice of $S$  and adjust the constant part of $a_D$ proportional 
to quark masses accordingly, as explained in \cite{KT}.
The final result for the physical quark number is the same, whatever 
initial choice for $S$ one makes, but the final formulas look most  
elegant with our choice. } 
The dynamical, fractional part of  quark number charges can be
determined as follows.

The structure of the low energy effective Lagrangian,
  when one of these monopoles
is  light (near one of the singularities in the $u$ plane),
is  then fixed  by
their {\it integer} quantum numbers, $n_m, n_e$ and 
\beq n_i \equiv S_i^{(cl)}, \quad i=1,2,\ldots N_f. \eeq
 Near the singularity of QMS  where a
$(n_m, n_e, \{n_i\})$ dyon is light $ L^{(light)}$   has the form
(assuming there is only
one such dyon)
 \bea
&& L^{(light)} =
\int d^4\theta \,[ M^{\dagger}e^{ n_m V_{0 D}+n_e V_{0}+\sum n_i V_{i} } M +
{\tilde M}^{\dagger}e^{-n_m V_{0 D} - n_e V_{0}- \sum n_i V_{i}}{\tilde M}]+
\non \\ &+&
\int d^2 \theta \, \sqrt{2}(n_m A_{0 D} +n_e A_0 +\sum n_i A_i){\tilde M}M
+ h.c.
\label{lagmono2}
\eea
The fact that the monopole  is coupled to the weak $U(1)$ gauge
fields with the (apparent) integer charges, does not mean that  its physical
charges are equal to the classical ones. The point is that there are
nontrivial boundary
effects to be taken into account, just as the
 Witten's effect  for the electric charge of the monopole, in the presence of
the $\theta$ term,  $(\theta /32 \pi^2) F_{\mu \nu} {\tilde  F}^{\mu \nu}$
\cite{Witten,Coleman}.

In our case, the crucial term is
the mixed gauge kinetic term, $\tau_{0i} W_0 W_i $ of
Eq.(\ref{lagrangianbis}). In fact,
this term yields a term   in the  energy
\beq    {1\over 4 \pi} \Re \,\tau_{0i} \int d^3x \,{\bf E}_i \cdot {\bf H}_0
\label{mixed}\eeq
where ${\bf E}_i$ and ${\bf H}_0$ stand respectively  for the
"electric" field associated with the weak, quark number $U_i(1)$ and for the
"magnetic" field associated with the strong $U_0(1) $ (related to the $SU(2)$)
gauge interactions.  In  the presence
of a static monopole,
  \beq   \int d^3x \,{\bf E}_i \cdot {\bf H}_0  \simeq
\int d^3x \, (-\nabla \phi_i(x)) \cdot \nabla {n_m  \o   r}=
 - 4 \pi n_m  \int d^3x
\, \phi_i(x) \delta^3({\bf x}),   \eeq
hence Eq.(\ref{mixed}) implies that the magnetic monopole, when observed at
spatial infinity, possesses an
additional quark number charge,
 \beq  \Delta S =n_m \, \Re \, \tau_{0i} = n_m \, \Re
{\de^2 F \o \de a \de a_i}. \eeq
The true, physical $i$-th quark number charge  of such a dyon is therefore
given by
\beq  S_i^{(phys)}= n_i+ n_m  \, \Re \, \tau_{0i}. \label{res}\eeq
This, which  generalizes   Witten's well-known formula \cite{Witten}, is
our main result.

\noindent{\bf  Generalization to \mbox{\boldmath $SU(N_c)$} }

The fractional quark numbers of dyons in $N=2$
supersymmetric $SU(N_c)$ gauge theories \cite{SUN} can be found  by   
similar  considerations. In the QMS the $SU(N_c)$ gauge group is broken 
to $U(1)^{N_c}$. The dyon carries  magnetic and 
electric charges of each unbroken $U(1)$ and the associated  quantum numbers
are denoted  by $(n_m^r, n_e^r), (r=1,\cdots,N_c)$. The gauge couplings
in the low energy effective action are also generalized to 
$\tau_{rs}$, $(r, s=1,\cdots,N_c)$, with  the
mixed $\theta$ terms, $(\theta_{rs}/32\pi^2)F_{\mu\nu}^r 
{\tilde F}^{s \mu\nu}$.
Therefore the $r$-th physical electric charge $Q^r$ is found to be
\beq
Q^r=n_e^r + \sum_{s=1}^{N_c}\mbox{Re}\, \tau_{rs}~n_m^s.
\eeq

In order to define  the physical quark numbers unambiguously we consider
$SU(N_c)\times U(1)^{N_f}$ gauge theories. The 
$U(1)^{N_c}\times U(1)^{N_f}$ effective Lagrangian is given by
\beq
L^{(eff)} = {1\over 8 \pi} \im \, \sum_{I,J} 
\int d^2 \th\,{1\o 2} \tau_{IJ}
W_I W_J + \sum_{I} \int d^4 \th
\,A_{D I} {\bar A}_I   + L^{(light)}, \label{lagrangiansun}
\eeq
where $I, J$ denote the combined suffix 
$(r,i), (r=1,\cdots,N_c, i=1,\cdots,N_f)$. From the mixed couplings
between the ``color'' $U(1)$ and the ``flavor'' $U(1)$ field strengths 
we find the physical $i$-th quark number of a dyon in $SU(N_c)$ gauge
theories as
\beq
S_i^{(phys)}=n_i + \sum_{r=1}^{N_c}\mbox{Re}\,\tau_{ri}~n_m^r.\label{ressun}
\eeq

\noindent{\bf Minimal coupling}

One might wonder whether, having an "exact low energy effective Lagrangian" at
hand,   such fractional quark number charges should not appear as  part of the
standard minimal interaction terms.  In fact, it is possible to
interpret  our result
this way.    In the case of
Witten's effect this was pointed out in \cite{DPK}.

There is indeed a large  class of arbitrariness in the choice of 
"dual" variables
$A_D \equiv A_{0D}, \, V_D \equiv V_{0D},$
corresponding  to a shift of these variables by terms  linear in 
$A,  \, \hbox{\rm and} \,
 A_{i},$  $  V,\, \hbox{\rm and} \,
V_{i},\,\,i= 1,2,\ldots ,  N_f$.  
These make a subgroup of $Sp(2+2N_f, R)$ which leaves 
Eq.(\ref{lagrangianbis}) form invariant (see however below).  
Actually, since the quark-number $U(1)$ groups are only weakly 
gauged, we can exclude those elements of
$Sp(2+2N_f, R)$ which transform 
\mbox{ $A_i \,\,(i=1,2,\ldots,  N_f)$}
 to their duals. In such a
case, the most general form of the relevant subgroup of $Sp(2+2N_f, R)$
have been
recently found by Alvarez-Gaum\'e et. al. \cite{LAZ};
they have the following general form: \beq  \pmatrix{  A_D \cr A \cr  A_{iD} \cr
A_i } \to    \pmatrix{ \alpha A_D + \beta A + p_i A_i  \cr  \gamma A_D +
\delta A
+ q_i A_i  \cr  A_{iD} + p_i(\gamma A_D + \delta A ) - q_i(\alpha A_D + \beta A)
-p_i q_i \cr A_i },\eeq
where $\pmatrix{\alpha & \beta  \cr \gamma & \delta } \in SL(2, R) $ and
$p_i, q_i$ are real.\footnote{In \cite{LAZ} models with
$N=2$ dilaton and mass "spurion" fields  are studied  and this leads them to
consider a  $Sp(4+2 N_f, R)$ group. Here we restrict ourselves to
renormalizable theories: this leaves  only the  quark masses to be  replaced by the
$N=2$ mass "spurion" fields in their language. The latter  is equivalent to
gauging
the $N_f$ quark-number $U(1)$ groups, as formulated here.    Note that we use a
slightly different notation from \cite{LAZ},  $a_i,\,a_{iD} $ instead of $m_i,\,
m_D^i$, etc. Note that $p_i, q_i$ are any {\it real} numbers, rational or
not.}

 The  transformations relevant to us are the ones  with
  $\pmatrix{\alpha & \beta  \cr \gamma & \delta } =
\pmatrix{1 & \beta  \cr 0  & 1 }$  and $p_i$.   These transformations leave the 
 effective Lagrangian Eq.(\ref{lagrangianbis}) invariant, {\it except for}
the shift, 
\beq \tau    \to \tau + \beta; \quad \tau_{0i} \to \tau_{0i} + p_i. \eeq
Therefore all of $\Re \, \tau$, $ \Re \, \tau_{0i}$  can be  eliminated by an
appropriate such transformation, i.e., by choosing
$\beta= \Re \, \tau, \,\, p_i= \Re \, \tau_{0i}.$  However 
 $N=2$ supersymmetry imposes a
simultaneous shift of vector superfields
\beq     V_D \to V_D + \beta V + p_i V_i ;\quad  V \to V;\quad
 V_{i} \to  V_{i}; \label{shiftgen}\eeq
in  the effective Lagrangian  involving light dyons such as
Eq.(\ref{lagmono2}).
The net result is that   the real part in the coefficients of the mixed kinetic
terms discussed  previously has disappeared and,  at the same time,
the dyon  with integer quantum numbers $(n_m, n_e,
n_i)$ is  now coupled minimally to the vector fields $A_{\mu}$ and  $A_{i
\mu}$   with charges
\beq   Q_e= n_e + n_m\, \Re \,\tau  ; \quad \hbox{\rm and
}\quad
  S_i^{(phys)}=  n_i + n_m \,  \Re \,\tau_{0i}.  \label{mainres}\eeq
The first is Witten's effect ($\Re \,\tau = \theta_{eff}/\pi$),  the second
is our result.

The  apparently  local effective  Lagrangian Eq.(\ref{lagrangianbis}) (with
 Eq.(\ref{lagmono2}))
with a nontrivial boundary effect,  has been transformed by
(\ref{shiftgen})  into an
explicitly nonlocal Lagrangian. Such an equivalence is to be expected after
all,  in view
of the dyonic nature of our monopoles.

\smallskip
Our formula for the physical quark number  Eq.(\ref{mainres}) can be
expressed in
terms of    known  quantities.   Note that in the limit of weak $U(1)^{N_f}$
couplings  the low energy
effective  coupling and $\theta$ parameter  of the $SU(2)$ sector as well as
the mass formula  must be the same as in \cite{SW2}. It means that the
prepotential is essentially that given in  \cite{SW2}:
\beq   F(a_0, \{a_i\}; \Lambda; \{\Lambda_i\}) =
 F^{(SW)}(a_0, \{m_i\}; \Lambda)|_{m_i=\sqrt2 a_i}  +
\sum_{i=1}^{N_f} C_i a_i^2,
\eeq
where  possible terms linear in  $a$ and $a_i$ have been dropped.\footnote{
The low energy effective action involves the second or higher derivatives of the
prepotential.  The term linear in  $a$  does affect the constant part of $a_D$
which should be fixed by appropriate convention so that the mass formula is
obeyed.}
The last term contains the genuine free
parameters of the theory, the $U_i(1)$ coupling constants at a given scale,
or the
position of the corresponding Landau poles, $\{\Lambda_i\}$.  Although these
affect the $U_i(1)$ coupling constants $g_i$,  they do not enter
 the calculation of the corresponding {\it charges},
 Eq.(\ref{res}).

\smallskip

$\tau_{0i}$ can now be expressed as
 \beq \tau_{0i} =  {\de^2 F \o \de a \de a_i}=  {\de a_D \o \de
{a_i}}\longvert_a = {\de  a_{iD} \o \de a}\longvert_{a_i},\quad
{\hbox{\rm where}} \quad
 a_D= {\de F \o \de a} = {\de F^{(SW)}(a, \{m_i\}; \Lambda) \o \de
a}. \eeq
The partial derivative of $a_D$ with respect to $a_i$  can be further
rewritten as
\beq      \tau_{0i} =   {\de a_D \o \de {a_i}}\longvert_a= {\de a_D \o \de
{a_i}}\longvert_u - {d a_D \o da} { \de a \o \de {a_i}}\longvert_u=
 {\de a_D \o \de {a_i}}\longvert_u -
\tau  { \de a \o \de {a_i}}\longvert_u. \label{formula1}   \eeq
The fractional quark charge can now be computed
  by using the known exact solution for
\beq {da_D\o du }=\oint_{\alpha} \omega,\quad {da\o du}=\oint_{\beta} \omega
,\quad     a_D=   \oint_{\alpha} \lambda_{SW},\quad
 a=\oint_{\beta} \lambda_{SW}, \eeq
and their derivatives with respect to $a_i= m_i/\sqrt2$.
The meromorphic differential   $\lambda_{SW}$, related to $\omega$ by
$ \omega=  {\sqrt2  \o 8\pi} {dx \o y}=   {\de \lambda_{SW} \o \de u}, $
is given explicitly in \cite{SW2,LAZ,KT,BF}.

\noindent{\bf Riemann bilinear relation}

 An  equivalent alternative formula for $\tau_{0i}$  can be found 
by  first rewriting
the formula (\ref{formula1})  as
\beq  \tau_{0i} =  \left( {\de a_D \o \de a_i}\longvert_u  {\de a \o \de
u}\longvert_{a_i}  - {\de a_D \o \de u}\longvert_{a_i} {\de a \o \de
{a_i}}\longvert_u
\right)/  {\de a \o \de u}\longvert_{a_i}. \eeq
In terms of a
meromorphic differential
$ \phi_i=  {\de \lambda_{SW}/\de a_i},$
and the holomorphic differential $\omega$, this can be written, 
 by using Riemann bilinear relation \cite{AG},
as
\beq   \tau_{0i} = \left( \oint_{\alpha}  \phi_i   \oint_{\beta} \omega -
\oint_{\alpha} \omega  \oint_{\beta}  \phi_i    \right)/ \oint_{\beta} \omega
=  2 \pi i \sum_{n} {\hbox {\rm
Res}}_{x_n^+}\phi_i \,\int_{x_n^-}^{x_n^+}  \omega / \oint_{\beta}
\omega, \label{formula2}
\eeq
where $x_n^+$ and $x_n^-$ denote the poles of $\phi_i$ in the first
and the second Riemann sheet respectively. The contour
 from $x_n^-$ to $x_n^+$ must be taken  so as to go around the
branch point which is not encircled  by
 the $\alpha$-cycle.  The positions of the poles 
$x_n$'s (which are nontrivial for $N_f=3$)
are explicitly given in \cite{LAZ}.

\noindent{\bf Semiclassical limit}

As a first  check of our result consider the semiclassical limit,
$u \gg \Lambda^2,\,\,
 a_i= m_i/\sqrt2 \gg \Lambda$.  In this limit Eq.(\ref{res}) must
reduce to the known
result
\cite{Ferrari,Niemi}:
\beq
S_i^{(phys)} \simeq n_i
+{n_m \o 2 \pi}\Arg {a + m_i \, /\sqrt{2} \o m_i /\sqrt{2} - a}.
\label{qnumbsc}
\eeq
 which is indeed the case as  can be
verified    by a direct calculation,  similar to the one in  the Appendix C
of
\cite{KT}. \footnote{In \cite{KT}  the present authors studied the electric
and quark number
fractionalization    in the context of the original $SU(2)$ gauge theory
with $N_f$ quark
hypermultiplets.  For the quark number charges,  it was only possible 
to make a check through the
mass formula, which
is known both semiclassically and exactly \cite{SW2}.}

\noindent{\bf Vanishing quark numbers of massless monopoles}

 An interesting special case is that of the
quark number charges of a massless
dyon, at one of the singularities of QMS.
 Since such an object occurs in  a theory
in which the original $SU(2)$  coupling constant becomes large in the
infrared,
the   semiclassical
method does not apply there.

For simplicity we consider the cases in which the singularity
is nondegenerate, with  only one massless monopole.
This is the case for $N_f=1$, or for $N_f=2$ or
$N_f=3$ with generic and nonvanishing masses.   {\it The physical quark
numbers of
these massless monopoles turn out to be zero.}

To see how this result comes about,  let us consider the one-flavor case
and concentrate on
the  $(n_m, n_e,  n_1)= (1,0,0)$  monopole occurring at the singularity
$u=u_3=  e^{-i \pi /3}$. (We use the unit,
$3 \cdot 2^{-8/3} \, \Lambda^2=1.$) The proof in other cases is similar.
From Eq.(\ref{formula1})  one has
\beq  \tau_{01} =   -{1 \o 2\pi} \left( \oint_{\alpha} {dx \o x y} - \tau \,
\oint_{\beta} {dx \o x y} \right); \quad   \tau= {d a_D \o da}=
 \oint_{\alpha} {dx \o y}   / \oint_{\beta} {dx \o y},  \eeq
where use of made of the explicit formulae valid for $N_f=1$,
$\lambda_{SW} = 
-(\sqrt2/4 \pi) y\, dx / x^2$; 
$y= x^2(x-u)+ 2 \sqrt2 a_1 x -1.$
Near $u \simeq u_3$,
two of the branch points  $x_2, \, x_3$ are close to each other
(and coalesce at $2 u_3/3$ when $u=u_3$), while the third one is near $-u_3/3$.
In the integrations over $\alpha$ cycle (which encircles the nearby branch
points
 $x_2$ and $x_3$)
$y$  can be approximated as
$ y =  \sqrt{(x-x_1)(x-x_2)(x-x_3)} \simeq    \sqrt{x_2- x_1}\,
\sqrt{(x-x_2)(x-x_3)},$
so that
\beq    \oint_{\alpha} {dx \o xy} = 2 \int_{x_3}^{x_2} {dx \o xy}  \to
 { 2 \pi \o  x_2  \sqrt{x_2- x_1}}, \qquad
\oint_{\alpha} {dx \o y} \to  { 2 \pi \o   \sqrt{x_2- x_1}},\eeq
as $u \to u_3$.
   On the other hand,   the integration over $\beta$ cycle is also dominated by
the region near  $x \simeq x_2$, and
\beq   \oint_{\beta} {dx \o xy} = 2 \int_{x_2}^{x_1} {dx \o x y}
 \simeq { 2  \o  x_2  \sqrt{x_2- x_1}} \cdot I, \eeq
where the integral $I=\int_{x_2}^{x_1} {dx \o
\sqrt{(x-x_2)(x-x_3)}}$  is divergent  at $u=u_3$.  However the $
\oint_{\beta} {dx \o
y}$ in the denominator of $\tau$ also diverges as
\beq  \oint_{\beta} {dx \o
y} \simeq    { 2  \o  \sqrt{x_2- x_1}} \cdot I. \eeq
Therefore
\beq     \tau_{01} \to   -\{{1  \o  x_2  \sqrt{x_2- x_1}} - {1  \o
x_2\sqrt{x_2- x_1}}
\cdot{
I \o I } \} =  0,\quad   S^{(phys)} \to 0,  \eeq
as $u \to u_3$.
The same result follows by using the second formula Eq.(\ref{formula2}).

For the massless $(1,1,1) $ dyon  at $u=u_1=  e^{+i \pi /3}$
(where the combined $\alpha + \beta$ cycle vanishes)
one finds by a similar analysis that $S^{(phys)}=1-1 =0.$

These results are somewhat analogous to the fact that all massless "dyons"
of $N=2$ Seiberg-Witten  theories with various $N_f$,   are actually all pure
magnetic monopoles with  $Q_e=0.$ \cite{KT}

\noindent{\bf Some numerical results}

It is straightforward to evaluate $\tau_{0i}$ given by the formula
(\ref{formula1}) or (\ref{formula2}) numerically at any point on the 
moduli space. In Fig.1 we show Re$(\tau_{01})$, or the physical
quark number of $(1,0,0)$ dyon in N=2 SQCD with a single
massless flavor. In this case $S^{(phys)}$  approaches  $-1/2$
in the weak coupling limit,  while it 
rapidly reduced to zero near the singularity  where the $(1,0,0)$ BPS state
becomes  massless, in accordance with the discussion in the precedent
paragraph.    The quark number of the same dyon
remains  equal to  $-1/2$    in  any vacua with real positive $u$.  

\begin{figure}[htb]
\begin{center}
\epsfxsize=0.6\textwidth
\leavevmode
\epsffile{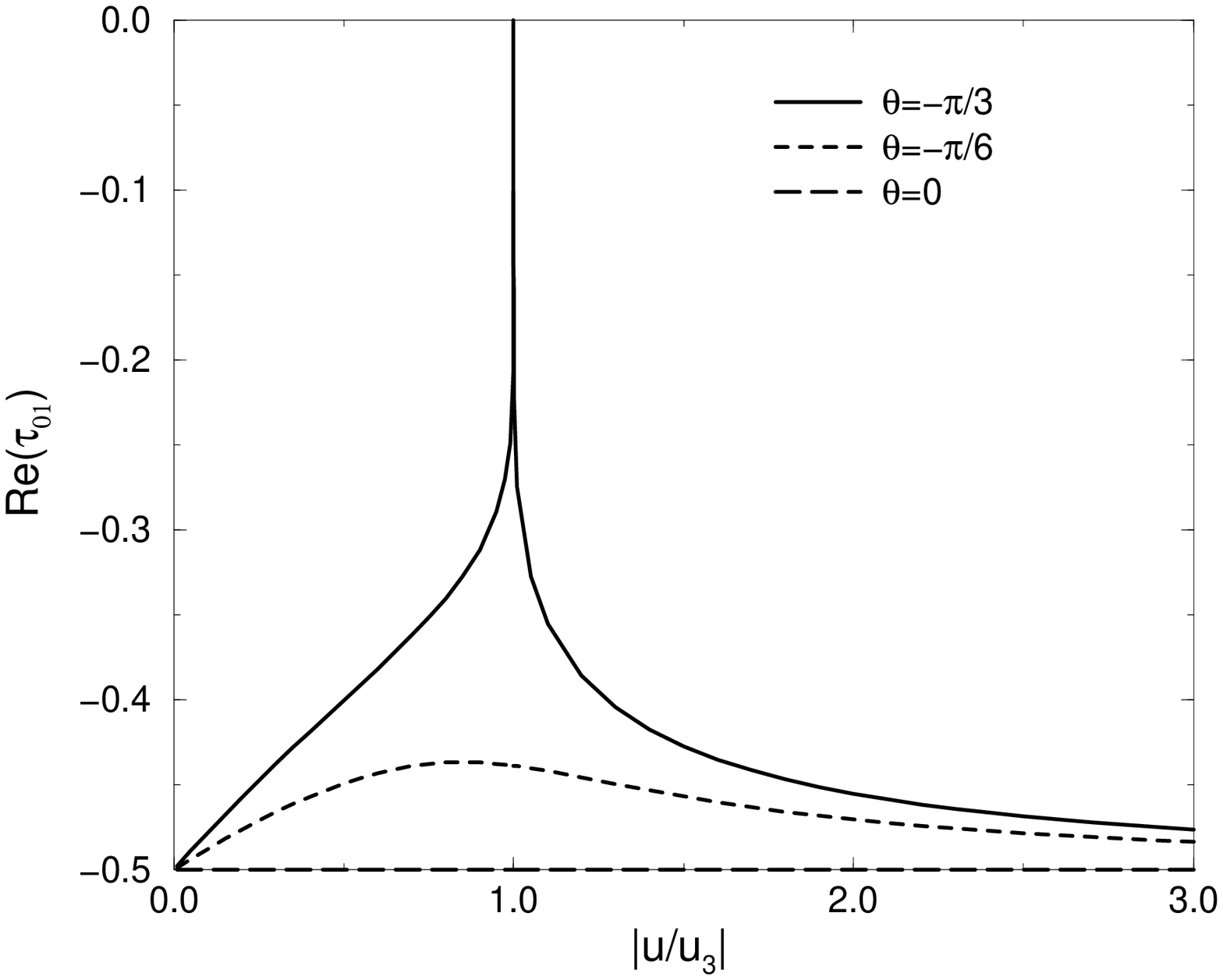}\\
\parbox{130mm}{
Fig.1: Re$(\tau_{01})$ in the massless $N_f=1$ theory is shown along
the half lines $u=|u|\exp{i\theta}$ of $\theta=0,-\pi/6,-\pi/3$. }
\end{center}
\end{figure}

\noindent{\bf CP invariance and quark numbers}

It is somewhat surprising  that the physical quark number of
the monopole takes all possible fractional values  even in a CP invariant
theory.
An example is  the $N_f=1$, $m=0$  theory at
$\Arg \, u = -\pi/3; \,\, |u| \ge  1 ,$  where $\theta_{eff}=0$.
Indeed $S^{(phys)}
$ of the $(1,0)$ monopole takes all real values from $0$  (at $u=
e^{-i \pi/3} $) to $-1/2$ (at
$|u| \to \infty$).  Such a result seems to be at odd with the
well-known result of Jackiw and Rebbi \cite{JR}, that
in a CP invariant $SU(2)$ theory
with a fermion in the fundamental representation the 't Hooft-Polyakov
monopole
becomes  a degenerate doublet with  fermion numbers $ q_{\pm}= \pm
1/2$.

The key for  solving  this apparent puzzle lies in the vacuum degeneracy.
In the argument of
Ref \cite{JR} the standard monopole $|0
\ket  \equiv  |M_-\ket$   is accompanied by another state
\beq    |M_+\ket = b^{\dagger} |M_-\ket, \quad {\hbox{\rm such that}}
\quad   \bra
M_-| \psi | M_+ \ket \ne 0  \label{M1}\eeq
where $b$ is  the fermion zero mode operator,
$ \psi = b\, \psi_0(x) + {\hbox {\rm nonzero}}\,{\hbox
{\rm modes}}$,
$  \{b, b^{\dagger} \}=1.  $  There is a  conserved fermion
conjugation symmetry  ${\cal F}$, such that
 \beq   [{\cal F}, H] =0;
\quad {\cal F} \, b \,{\cal F} =b^{\dagger}, \quad {\cal F}^2 = {\bf
1}.  \label{fermnum}\eeq
 This last equation, together with
Eq.(\ref{M1}), implies that the  monopole states $|M_{\pm} \ket$
transform to each other by ${\cal F}$:
$  {\cal F} |M_{\pm} \ket = |M_{\mp} \ket. $
The fermion number operator must be defined as
\beq   S= {1 \o 2}  \int d^4x \,  ( \psi^{\dagger} \psi -
\psi  \psi^{\dagger} ) = {1 \o 2} ( b^{\dagger} b - b b^{\dagger} ) +
\ldots    \eeq
so  that
\beq  {\cal F} S  {\cal F} = -S, \quad   [S, \psi]= -\psi.
\label{ingred1}\eeq
One finds
  $q_+=-q_-$ from the first of the above.    On the other hand, from the
$\bra M_-| \ldots
|M_+ \ket$ matrix element of the second of  Eq.(\ref{ingred1})
another relation, $ q_+= q_- +1$ follows.   Combining these two  one
finds  the announced result  $q_+= -q_-=1/2$.  Of course, the first of
Eq.(\ref{fermnum}) guarantees that the states  $|M_{\pm} \ket$
are degenerate in mass.

In the present theory the role of the fermion conjugation is played by CP
symmetry. Under CP, however, the vacuum is also transformed as
$  u \to u^{*}. $
What happens in the case of theories at
$\Arg \, u = -\pi/3; \,\, |u| \ge 1,$  is that the theory is transformed by
CP to another theory, related to the original one by an exact $Z_3$ symmetry.
This explains the fact that  $\theta_{eff}=0$ and that the low energy
effective monopole theory there has an exact CP invariance in the usual
sense \cite{KT}.   Nonetheless, from the formal point of view the original CP
symmetry  of the underlying theory  is
spontaneously broken in this case, and the Jackiw-Rebbi argument does not apply.
In fact, although the operator relations Eq.(\ref{fermnum}) and
Eq.(\ref{ingred1}) still hold, the states are now transformed by
$  {\cal F} |M_{\pm}; u \ket = |M_{\mp}; u^{*} \ket:$  
the two states related by CP operation live on two different Hilbert space.
As a result,  the first of Eq.(\ref{ingred1}) yields
$  q_+^{*} = - q_-; \quad  q_+ = - q_-^{*}, $
where $q_{\pm}^{*}$ are the quark  number of  the states $|M_{\pm}; u^{*}
\ket$.
The second equation, whose matrix elements relates the states in the same
vacuum,  leads to
$  q_-= q_+ -1; \quad   q_-^{*}= q_+^{*}  -1. $
Note that these four relations are mutually consistent  and relates the four
charges by $q_+^{*}  = - q_- = -q_+ +1 = q_-^{*} +1. $
Although the first of Eq.(\ref{fermnum}) does imply that the monopoles
$|M_{+}; u^{*} \ket$ and $|M_{-}; u \ket$ have the same mass,
(and similarly $|M_{-}; u^{*} \ket$ and $|M_{+}; u \ket$)
it does not imply any degeneracy;  it rather  means that
 the spectrum of the theories  at $u$
and at $u^{*}$ are the same, reflecting the
  $Z_3$ symmetry of the underlying theory .

Note that along the real positive values of $u$ (for $N_f=1$),  where CP is
exact and not spontaneously broken (with a CP invariant vacuum),
dyons are found indeed to be doubly degenerate
and have quark numbers $\pm 1/2$, in accordance
with  \cite{JR}.

\smallskip

\noindent{\bf Quark-number current correlation functions}

Once  the physical quark numbers of light dyons are known, the analogue of
the R-ratio
associated with  the correlation function Eq.(\ref{cortrel}) may be
computed at low energies
by the one loop contributions of the weakly coupled dyons. By an
appropriate normalization
one finds that, near a nondegenerate singularity,  the light monopoles and the
fermion partners
$\psi_M, \,{\tilde \psi}_M, \, M, {\tilde M}, $ add up to give
  $Disc_{Q^2}\Pi(Q^2) \simeq 3 \, (S^{(phys)})^2,$ for  $Q^2 \ll \Lambda^2,
$  while  at high
energies
 quarks and squarks yield      $ Disc_{Q^2}\Pi(Q^2) \simeq
N_c \,  (1 + 1 + 2 \cdot (1/2) ) = 6. $

\bigskip
\noindent{\bf Acknowledgement}
 One of the authors  (K.K.) thanks 
Yukawa Institute  of Theoretical Physics, Kyoto University,  where part of this  work
was  done,   for a warm hospitality. He also  wishes to express his gratitude to Japan
Society of Promotion of Science for a generous support in the context of Visiting 
Research Project. H.T. thanks the Department of Physics,
University of Genova for hospitality and INFN for
financial support during his stay at Genova. 
He is also supported in part by Grant-in Aid for Scientific
Reserach (\#08640361) from the Ministry of Education,
Science and Culture of Japan.   G.C. thanks 
N. Magnoli for many useful discussions.

\end{document}